\documentclass{wnna}
\usepackage{natbib}\usepackage{txfonts}\usepackage{balance}
\usepackage{graphicx}
\usepackage[a4paper]{hyperref}
\idline {1}{101}

\begin{document}

\title{The ignition process in type Ia supernovae: numerical simulations of core temperature perturbations}

\author{
L. \,Iapichino \inst{1} \and M. \,Br\"uggen \inst{2} \and W. \,Hillebrandt \inst{3} \and J.C. \,Niemeyer \inst{1}
          }

  \authoremail{luigi@astro.uni-wuerzburg.de}

\institute{Lehrstuhl f\"ur Astronomie, Universit\"at W\"urzburg, Am Hubland, D-97074 W\"urzburg, Germany \and Jacobs University Bremen, Campus Ring 1, D-28759 Bremen, Germany \and Max-Planck-Institut f\"ur Astrophysik, Karl-Schwarzschild-Str. 1, D-85741 Garching, Germany}

\authorrunning{Iapichino L. et al.}
\titlerunning{The ignition process in type Ia supernovae}

\abstract{The onset of the thermonuclear runaway in a Chandrasekhar-mass white dwarf, leading to the explosion as a type Ia supernova, is studied with hydrodynamical simulations. We investigate the evolution of temperature fluctuations (``bubbles'') in the WD's convective core by means of 2D numerical simulations. We show how the occurrence of the thermonuclear runaway depends on various bubble parameters. The relevance of the progenitor's composition for the ignition process is also discussed.

 \keywords{Supernovae: general -- Hydrodynamics -- Methods: numerical
-- White dwarfs}

 }


\maketitle{
}

\section{Introduction}
\label{intro}

The transition from the hydrostatic to the explosive C-burning in a carbon-oxygen white dwarf (CO WD), which accretes mass in a binary system until it approaches the Chan\-dra\-se\-khar mass, marks the start of the explosion of this progenitor star as a type Ia supernova (SN Ia). From a computational point of view, the modelling of this phase is extraordinarily challenging because of the span of involved length and time scales at ignition. 

Nonetheless there are several studies on the evolution of the progenitor to the ignition of SN Ia explosion. \citet{lht06} study the accretion phase of the progenitor and find interesting links between binary population synthesis and the physical conditions of the WD at ignition. On shorter time intervals, the last stage before the thermonuclear runaway has been explored with analytical models \citep{wwk04,ww04} and numerical simulations \citep{hs02,kwg06}. Though widely different in their numerical approach, these studies agree in identifying the convective flow prior to the runaway as a crucial issue for the ignition process and the early stage of the explosion.

A complementary way to address the ignition problem is given by small-scale (length scales of the order of $1$\ km or less, whereas the WD diameter is about $2000$\ km) investigations of the evolution to runaway (e.g.~\citealt{gsb05}, \citealt{zd07}). According to the models of \citet{gsw95} and \citet{ww04}, the thermonuclear runaway is triggered by temperature perturbations (``bubbles'') on the km-scale, generated in the WD core by the turbulent convection. The study of the bubble features is therefore a powerful tool to better understand the ignition process.

In this contribution, the numerical simulations (Sect.~\ref{setup} and \ref{simulations}) of \citet{ibh06} (henceforth I06) are reviewed (we refer to that paper for further details). In Sect.~\ref{discussion} we extend the discussion of the results to assess the role of the WD composition and the relevance of new nuclear measurements \citep{srr07}.

\section{Numerical setup}
\label{setup}

The 2D simulations were performed using the FLASH code \citep{for00}. The bubble is initialised as a temperature perturbation, hotter and less dense than the background, and in pressure equilibrium with the surrounding matter. The size of the computational domain is $5 \times 20$\ km, the geometry is Cartesian. The background WD model (provided by S.~Woosley) is mapped in plane-parallel approximation, applying the procedure described by \citet{zdz02} to enforce the hydrostatic equilibrium. A constant gravitational acceleration, computed from the WD model, is applied to the domain, with the gravitational acceleration pointing downwards along the $y$-axis. 

The adaptive mesh refinement (AMR) is used, allowing five levels of refinement with an effective grid size of $[256 \times 1024]$ zones, corresponding to an effective spatial resolution of $2 \times 10^3$\ cm.  

A Helmholtz EOS \citep{ts00} is used. A minimal $\alpha$-network with 7 isotopes \citep{thw00} is implemented to follow the hydrostatic C-burning.  

\begin{figure}[t]
\resizebox{\hsize}{!}{\includegraphics[clip=true]{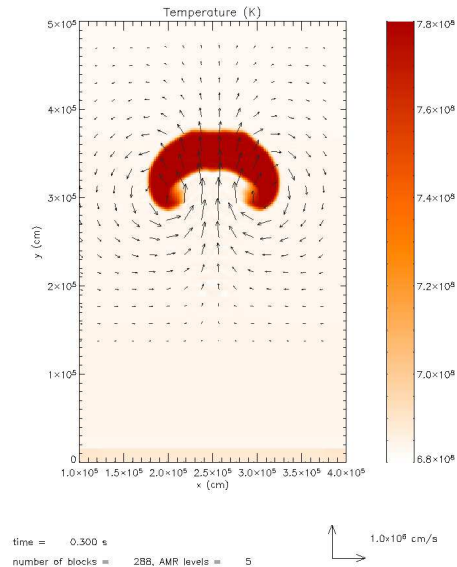}}
\caption{ \footnotesize Temperature plot of the rising bubble at $t = 0.3$\ s, with the velocity field superimposed, in a run with the initial parameters $T = 7.7 \times 10^7$\ K, $D = 1$\ km, $R = 100$\ km.} \label{vectors}
\end{figure}

As inferred by the analytical models \citep{ww04,wwk04}, bubbles in the WD core are generated with some range of sizes, temperatures and distances from the WD centre.  A series of thirteen calculations was performed over the range of expected values for these relevant bubble parameters i.e. the initial bubble temperature $T$ ($7.3-7.9 \times 10^8$\ K), the initial bubble diameter $D$ ($0.2-5$\ km) and the initial distance from the WD centre $R$ ($50-150$\ km). A snapshot of the bubble motion is shown in Fig.~\ref{vectors} (see I06 for the complete temporal sequence). 

\section{The physics of buoyant bubbles}
\label{simulations}

The evolution of the bubble is governed by the interplay between nuclear burning and hydrodynamical instabilities. The hydrostatic carbon burning determines the nuclear timescale, defined as the time needed by the temperature perturbation to reach the thermonuclear runaway ($T \sim 10^9$\ K), of the order of some seconds. On the other hand, the bubble is subject to the Rayleigh-Taylor instability. It is accelerated upwards by the effective gravitational acceleration $g_{\mathrm{eff}}(r) = g(r)\, \Delta \rho / \rho$, where $g$ is the gravitational acceleration at the distance $r$ from the WD centre and $\Delta \rho / \rho$ is the density contrast between the bubble and the background. The vortical motions produced by the bubble's buoyant rise, shown in Fig.~\ref{vectors}, lead to its fragmentation and dispersion down to length scales in the range $10 - 100$\ cm, where the heat loss by thermal conduction is larger than the nuclear heating. The dispersion timescale is well approximated by the buoyant rise time $\tau_{\mathrm{brt}} = D / v_{\mathrm{b}}$, where $D$ is the bubble diameter and $v_{\mathrm{b}}$ is the terminal bubble velocity (of the order of $1\ \mathrm{km\ s^{-1}}$).

\begin{figure}[t]
\resizebox{\hsize}{!}{\includegraphics[clip=true]{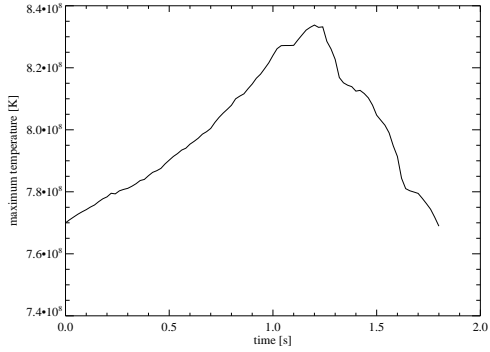}}
\caption{ \footnotesize Evolution of the maximum bubble temperature in a simulation with the initial parameters as in Fig.~\ref{vectors}. } \label{maxtemp}
\end{figure}

The outcome of the bubble evolution is determined by these competing effects and, quantitatively, by the corresponding timescales. If the nuclear timescale is smaller than the buoyant rise time, the bubble reaches the thermonuclear runaway before being dispersed. Conversely, the bubble is dispersed and cools down before a thermonuclear flame is ignited. The evolution of the bubble temperature in the latter case is shown in Fig.~\ref{maxtemp}, where $T$ increases until the dispersion prevails. Further numerical issues in this process and limitations in our approach are discussed in I06.

\section{Results and discussion}
\label{discussion}

\subsection{The parameter study and its implications for SN Ia ignition}

The role of the bubble parameters on the physics of temperature perturbations has been explored with the performed parameter study. The main results are reported here:

\begin{itemize}
\item The bubble diameter $D$ affects the buoyant rise time because $\tau_{\mathrm{brt}} \propto D^{1/2}$ ($v_{\mathrm{b}} \propto D^{1/2}$ as well, as in \citealt{g02}). Larger bubbles have longer $\tau_{\mathrm{brt}}$ and are therefore favoured to go to thermonuclear runaway. 

\item The bubble temperature $T$ affects the nuclear timescale, because of the high sensitivity of the burning to $T$. Hotter bubbles are favoured to go to runaway. However, to some extent, also the buoyant rise time decreases with increasing temperature, because a larger temperature contrast leads to an increase of  $g_{\mathrm{eff}}$, producing a more effective bubble dispersion.

\item The increase of the central distance $R$ increases $g_{\mathrm{eff}}$ and thus decreases the buoyant rise time, because both the bubble temperature contrast and $g(R)$ increase with increasing $R$. Bubbles at smaller distance from the WD centre are favoured to go to runaway. 
\end{itemize}

The evolution of 2D bubbles is a highly simplified model for the small-scale ignition, nonetheless the results can be used to provide some clues on the ignition process in SNe Ia at larger length scales. First, we observe that the estimated convective velocity before the runaway ($50 - 100\ \mathrm{km\ s^{-1}}$) is much larger than the bubble buoyant velocity. The features of the convective flow in the progenitor prior to runaway are therefore needed to model the ignition properly. Comparing convective velocities and timescales for bubble evolution, we estimate that the ignition is initiated at a central distance of about $150$\ km, with a dispersion of the order of $100$\ km, which agrees well with previous results \citep{wwk04,ww04}.

The bubbles are generated in the WD's core with some probability distribution function of temperature fluctuations. Our bubble analysis indicates that the (few) hotter bubbles are more effectively dispersed than the (many) milder temperature fluctuations, because of their shorter bubble rise time. It suggests, indirectly, that the multi-point ignition scenario is a viable model for the onset of the SN Ia explosion.   

\subsection{Role of other parameters}

The study of the diversity in explosion simulations (e.g.~\citealt{rgr06}), caused by the physical parameters of the WD and linked with the observed diversity in SNe Ia, has received comparatively more attention than the exploration of the diversity in the ignition phase. The bubble model described here has intrinsic and numerical limitations in its predictive power of the ignition properties of SNe Ia, but it can be profitably used as a ``probe'' of the influence on the ignition process of other physical parameters. 

In \citet{rgr06} the role of the WD composition is explored by varying the carbon/oxygen ratio and $^{22}\mathrm{Ne}$ mass fraction (representative of the progenitor's metallicity, \citealt{tbt03}). \citet{dt06} found that the ignition timescale depends strongly on the $^{22}\mathrm{Ne}$ abundance. We run some test simulations which confirm their results in the framework of the bubble physics. We suggest that the importance of both the C/O ratio and $X(^{22}\mathrm{Ne})$ for the {\em ignition} process should be carefully scrutinised in future works. 
  
Recently \citet{srr07} presented new measurements of the cross sections of the reactions $^{12}\mathrm{C}(^{12}\mathrm{C},\alpha)^{20}\mathrm{Ne}$ and $^{12}\mathrm{C}(^{12}\mathrm{C},\mathrm{p})^{23}\mathrm{Na}$ from $E = 2.10$\ to $4.75$\ MeV. The authors claim that the new reaction rates have interesting astrophysical applications. The improvement of the experimental determination of these reaction rates in the considered energy range (or at lower energies) could significantly influence the hydrostatic carbon burning in the progenitor and is worth being explored. The resonant screening effects should also be further investigated \citep{itw03}.

\begin{acknowledgements}
The FLASH code is developed by the DOE-supported ASC / Alliance Center
for Astrophysical Thermonuclear Flashes at the Uni\-ver\-si\-ty of
Chicago. L.I. acknowledges the INAF -- Catania Astrophysical Observatory for the travel financial support for this workshop. 
\end{acknowledgements}


\bibliography{snia-ref} 
\bibliographystyle{aa}

\end{document}